# Microtransit adoption in the wake of the COVID-19 pandemic: evidence from a choice experiment with transit and car commuters

**Jason Soria**
PhD Candidate
Department of Civil and Environmental Engineering
Northwestern University
2145 Sheridan Road, Evanston, IL 60208, USA
Email: jason.soria@u.northwestern.edu

**Shelly Etzioni**
PhD Student
Dept. of Civil and Environmental Engineering
Technion, Israel Institute of Technology
Technion city, Haifa 32000, Israel
Email: shellybz@campus.technion.ac.il

**Yoram Shiftan**
Professor
Dept. of Civil and Environmental Engineering
Technion, Israel Institute of Technology
Technion city, Haifa 32000, Israel
Email: shiftan@technion.ac.il

**Amanda Stathopoulos**
Associate Professor
William Patterson Junior Professorship Chair
Department of Civil and Environmental Engineering
Northwestern University
2145 Sheridan Road, Evanston, IL 60208, USA
Email: a-stathopoulos@northwestern.edu

**Eran Ben-Elia**
Associate Professor
Dept. of Geography and Environmental Development
Ben-Gurion University of the Negev
PO Box 653, Beer Sheva, 8410501, Israel.
Email: benelia@bgu.ac.il

ABSTRACT

On-demand mobility platforms play an increasingly important role in urban mobility systems. Impacts are still debated, as these platforms supply personalized and optimized services, while also contributing to existing sustainability challenges. Recently, microtransit services have emerged, promising to combine advantages of pooled on-demand rides with more sustainable fixed-route public transit services. Specifically, microtransit can provide both dynamic rider-driver matching to serve demand with fewer vehicles, as well as designing optimal routes if riders accept to wait or board vehicles in curbside boarding locations. Understanding traveler behavior becomes a primary focus to analyze adoption likelihood and perceptions of different microtransit attributes. The COVID-19 pandemic context adds an additional layer of complexity to analyzing mobility innovation acceptance. This study investigates the potential demand for microtransit options against the background of the pandemic. We use a stated choice experiment to study the decision-making of Israeli public transit and car commuters when offered to use novel microtransit options (sedan vs. passenger van). We investigate the tradeoffs related to traditional fare and travel time attributes, along with microtransit features; namely walking time to pickup location, vehicle sharing, waiting time, minimum advanced reservation time, and shelter at designated boarding locations. Additionally, we analyze two latent constructs: attitudes towards sharing, as well as experiences and risk-perceptions related to the COVID-19 pandemic. We develop Integrated Choice and Latent Variable models to compare the two commuter groups in terms of the likelihood to switch to microtransit, attribute trade-offs, sharing preferences and pandemic impacts. The results reveal high elasticities of several time and COVID effects for car commuters compared to relative insensitivity of transit commuters to the risk of COVID contraction. Moreover, for car commuters, those with strong sharing identities were more likely to be comfortable in COVID risk situations, and to accept microtransit. We discuss the implications of the differences between these commuter groups.

# 1  Introduction

A sound public transit system, along with on-demand and shared forms of mobility, play a significant role in supporting the economic functioning and well-being of cities and their residents. Among mobility on demand services, smartphone-sourced ridehailing which uses online platforms for booking, payment, and communication to match drivers with riders, along with more recent ridepooling, which is ridehailing with multiple parties play an increasingly important role in worldwide urban mobility (Shaheen and Cohen, 2018). These services can complement transit and benefit both passengers and cities by improving accessibility while reducing transportation externalities such as air pollution and traffic congestion. Concurrently, there is increasing evidence from North American cities (New York, San Francisco, Chicago, Los Angeles, and Seattle) that ridehailing is a major contributor to traffic congestion (Graehler et al., 2019, Erhardt et al., 2021, Wu and MacKenzie, 2021) and may compete with mass transit (Yan et al., 2020). Measures like promoting ridepooling can help curb vehicle miles traveled (VMT), a negative impact of ridehailing. Yet, the effectiveness of ridepooling in reducing congestion has come under scrutiny and depends on deadheading and local pooling rates (Schaller, 2021)

In response to these concerns, transit agencies and mobility startups have launched microtransit services—small-scale, on-demand transit fleets that can offer both fixed routes and scheduled operations, as well as more flexible routes and on-demand scheduling (APTA, 2021). This new service model may produce environmental *and* rider benefits. It relies on information and communication technology (ICT) platforms to enable on-demand service requests or coordination between riders and drivers for trip pooling. This coordination makes the transition from door-to-door to curb-to-curb (e.g. at transit stops) services easier to implement.

The shift to microtransit calls for research on user behavior, motivations, and acceptability to understand demand and its impacts on mobility systems. Beyond traditional transit attributes like travel time and fare, microtransit entails new attributes related to curb-to-curb routing, scheduling, and different sharing configurations. Pinpointing how customers evaluate these new service dimensions is critical for researchers and decision-makers to design new mobility platforms complementing existing transportation systems. Different mode experiences are also likely to lead to different service feature perceptions. An attribute such as expected walking time to the boarding location can be viewed as a disadvantage against the baseline of private car or ridehailing but is a familiar factor for transit users. Understanding how riders weigh microtransit attributes is key to designing and maintaining an efficient transportation service. Platform managers, either from the public or private sectors, can analyze this demand to optimize their fleet, attract patronage and minimize passenger delays. What is more, knowledge of acceptability and attribute tradeoffs informs this mode's relationship to traditional commute options like personal vehicles and public transit, the outlook of public-private partnerships, and the need for additional infrastructure to support microtransit options (Shaheen et al., 2020).

During 2020 and 2021, the COVID-19 pandemic and associated restrictive measures have drastically disrupted mobility systems worldwide — adding an additional layer of uncertainty to mobility demand analysis. Stay-at-home orders, move to telework, and other social distancing measures to prevent the spread of the coronavirus, have led to steeply falling demand for mobility, especially public transit and shared vehicle mobility (Liu et al., 2020, Duarte, 2020, Higgins and Olson, 2020). Due to these changes and the lingering safety perceptions, the pandemic has likely heightened travelers' sensitivity to close physical interactions and consequently changed riders' priorities when trading off cost and comfort against health and safety. In 2021, as workers increasingly return to work, immunization rates increase, and people start commuting anew, the need for shared mobility services is growing. Therefore, the need to understand the links between pandemic risk perceptions and mode preferences remains an urgent research priority (Hensher, 2020). Yet, we have limited insight into how people navigate the decisions of using different types of shared modes during the evolving pandemic (Shokouhyar et al., 2021).

This research aims to analyze commuting travelers' acceptability of novel microtransit commute options in the wake of the COVID-19 pandemic. We address three specific research questions: *First*, we analyze user acceptance of microtransit options, emphasizing several new microtransit-specific attributes. Specifically, we examine the factors that explain the shift from the status quo commute to microtransit travel and analyze attribute sensitivities and elasticities. *Second,* we explore the differences between current transit users and solo drivers. Given that microtransit combines on-demand rides and mass transit services' features, we expect differences based on current commute modes. *Third*, we assess the joint impact of COVID experiences and concerns along with shared mobility and intrinsic motivations for sharing to build a new understanding of how vehicle pooling and other novel attributes are perceived in the COVID-19 context. Thereby, the evolving perception and potential recovery of shared mobility and the trade-offs between traditional and novel mode attributes are further elucidated. Additionally, our examination of pandemic perceptions and sharing experiences allows us to disentangle how different commuter groups view these novel services and attributes.

We use data from a web-based survey conducted in Israel following the first COVID lockdown in May 2020. The study included a choice experiment (CE) with a pivoted Bayesian efficient design. The CE scenarios present two microtransit alternatives using the respondents' status quo mode and their stated travel time and cost. The first is ridepooling in a sedan-sized vehicle with a passenger capacity of 4 (not including the driver) which we will refer to as Microtransit Sedan (MT-S). This service has not been introduced in Israel so far due to regulatory limitations. The second is ridepooling in a van-sized vehicle with a capacity of 10 passengers, which we will refer to as Microtransit Van (MT-V). This service is operated only on a limited scale—on a pilot basis in the main cities of Tel Aviv, Jerusalem, and Haifa and one rural area. Data about the respondents' sociodemographics, political views, COVID-19 attitudes, and sharing experiences is also collected. We employ an Integrated Choice and Latent Variable (ICLV) framework to examine the acceptance of these new commute options and the impact of user profiles, latent attributes of sharing motivations, and COVID perceptions.

Our analysis reveals three key takeaways. (1) New mode attributes significantly affect the utility of the microtransit alternatives, with a notable aversion to walking and waiting among drivers. (2) car and transit commuters have structural differences in attribute elasticities. (3) significant differences are noted for the magnitude of latent variable effects. For drivers evaluating microtransit, sharing experience and COVID Comfort play a key role in the decision-making. Overall, these results suggest that car commuters find out-of-vehicle travel and planning ahead highly unattractive. Transit users are much less affected by sharing and COVID constructs. We discuss the extent to which these results are due to captive transit users and the implications on their willingness to use microtransit modes for their commute.

The rest of the paper is organized as follows: We complete a literature review on on-demand mobility, focusing on microtransit research and the effects of COVID-19 on (shared mode) mobility. We then explain the survey instrument and data used in the analysis. In the next section, our methodology is defined. Finally, we discuss the results, implications, and conclusions to complete the analysis.

# 2 Literature Review

## 2.1 On-demand shared ride mobility and traditional mobility options

The increased presence and penetration of ridehailing platforms in urban mobility systems presents both opportunities and challenges for policymakers and planners. Ridehailing provides several benefits for users. These benefits include serving as a transit gap-filler, first-mile-last-mile connection to transit (Brown, 2018, Shaheen and Cohen, 2018), improving mobility accessibility for underserved communities (Brown, 2019), providing more personalized door-to-door services at lower fares than the traditional on-demand travel offered by taxis (Rayle et al., 2016), and avoiding limited parking (Clark and Brown, 2021). These benefits, however, may come at a cost.

Increasingly, negative externalities related to novel mobility platforms have been highlighted. Owing to deadheading and induced trip-making, ridehailing has been observed to increase VMT, congestion, and pollution (Graehler et al., 2019, Erhardt et al., 2019, Nair et al., 2020), as well as indications of possible negative impacts on transit ridership and its financial sustainability (Schaller, 2021). Several studies suggest that a large segment of users may substitute transit for hailing (Clewlow and Mishra, 2017, Dong, 2020). Aggregate trip data analysis similarly shows that demand for on-demand hailing is often higher where transit demand is also high (Brown, 2019, Correa et al., 2017). The demand relationship, and degree of substitution, vary according to several factors, such as transit service quality/coverage and type (Gehrke et al., 2019), and city geography and socio-economic factors (Jain et al., 2017, Hall et al., 2018, Soria and Stathopoulos, 2021).

Potential strategies to address negative mobility externalities associated with ridehailing are to promote increased ridepooling and adopt more features of public transit (which is discussed further in the following subsection). Pooled ride services have been offered by leading Transportation Network Companies (TNCs)— UberPool or Lyft Line—since 2014. Rather than one single vehicle exclusively serving a single rider request, multiple trips can be pooled together in the same vehicle to increase vehicle occupancy rates and reduce excess VMT (Hou et al., 2020). Cities can also play a part in shifting ridehailing demand towards more sustainable usage. For example, Chicago introduced an extra fee for single-party rides beginning or ending within the downtown area to encourage pooling (Pratt et al., 2019). In terms of behavior, research shows that cost considerations are still crucial in pooled rides (Morales Sarriera et al., 2017, Soria et al., 2020, Lavieri and Bhat, 2019). Other avenues to increase pooled rides are incorporating locally aligned values and culture into the sharing platform (Rong et al., 2021).

## 2.2 Emergence of Microtransit Options

Microtransit can be defined as a digitally-enabled transit service that is privately or publicly operated, using pooled shuttles or vans, to provide on-demand or fixed-schedule services with either dynamic or fixed routing (SAE, 2018). Calderón and Miller (2020) highlight the range of service types within microtransit. This service model is positioned between current (typically single occupancy) ridehailing, and traditional fixed-route transit, owing to the promotion of pooling rides, walking to the curb to connect with optimal routes, and scheduling rides in advance of boarding time. In practical terms, we can characterize microtransit as a new form of ridehailing with transit-like attributes that aim to optimize trips collectively by minimizing vehicle miles traveled. **Figure *1*** compares door-to-door ride-pooling with curb-to-curb microtransit. Fewer vehicles are needed to serve demand by pooling trips, thus reducing VMT (Fu and Chow, 2021). Providing curb-to-curb services where passengers walk to a designated boarding location and alight nearby their final destination reduces the amount of vehicle travel. In **Figure *1***, Party 1 (P1) and Party 2 (P2) can be served directly at their origins and destinations or meet the driver at a designated boarding location and alight nearby their destination. In the latter scheme, the vehicle travels less. Additionally, reserving a seat in a shared vehicle well ahead of boarding allows the operator time to pool trips optimally rather than relying on real-time driver-rider matching.

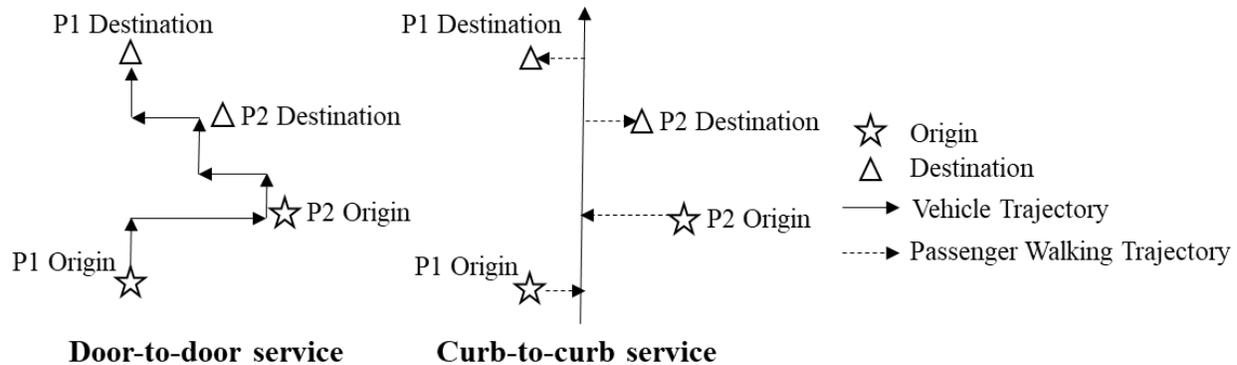

**Figure 1 Door-to-door and Curb-to-curb Service Types**

Research on van-based and other microtransit oriented ridehailing services is still limited. We can gain initial insight into the acceptance and behavior of shared rides by drawing on ridepooling and related literature. Large fleets of shared-taxis have been shown in simulations to serve existing taxi demand without excessively long delays, without significantly reducing revenue, and importantly, reducing VMT (Alonso-Mora et al., 2017, Martinez et al., 2015). However, achieving these outcomes requires a high market share of pooled trips, as VMT benefits can only be achieved with sufficiently large shared vehicles fleet sizes and passengers' demand (Rodier et al., 2016; Fagnant and Kockelman, 2018).

In reality, reported rates of ridepooling are typically low and vary considerably. Empirical estimates range from 6-35% (California Air Resource Board, 2019, Chen et al., 2018, Chicago Metropolitan Agency for Planning, 2019, Li et al., 2019, Lyft, 2018, Soria et al., 2020, Young et al., 2020). Lastly, there is much room to grow in terms of wider adoption of pooling in many cities, e.g. 94% of ridepooling trips in LA are made by just 10% of riders (Brown, 2020).

Few empirical studies are available to evaluate microtransit in practice. The "Breng flex" pilot in the Netherlands stresses the risk of an excess shift of users away from transit towards microtransit in response to pricing differences (Alonso-González et al., 2018). A study of three U.S microtransit pilots concluded that implementation was fraught, and low ridership was a recurring problem (Westervelt et al., 2018). Additionally, an Uber-based microtransit service case study found that it did not attract single-occupant vehicle users and instead mainly drew users away from public transit (Lewis and MacKenzie, 2017). Similarly, pooling users are found to typically be multimodal already (Kostorz et al., 2021). Lastly, microtransit users may highly enjoy the service but be unwilling to pay higher fees. The Finnish pilot Kutsuplus found that substantial subsidies were needed for the program to be financially viable (Rissanen, 2016). In sum, for shared microtransit services to succeed in mitigating externalities, further research is needed to understand the tradeoffs travelers are willing to make to pool rides and optimal service implementation.

## 2.3 Choice Experiment Analysis of Microtransit Features

Stated choice research offers valuable insight on demand related to microtransit and specific service features. In the following, we examine SC-based studies that analyze multiple microtransit attributes. The real-time matching and routing capabilities are central for expanding the transit-like aspects of ridehailing to inform pooling, and curb-to-curb routing, since new trip attributes can be communicated with potential riders. Therefore, SC surveys are essential tools to measure how riders view these new attributes, such as walking, pooling people, and additional passenger pickups. **Table *1*** summarizes relevant CE-based studies involving microtransit and related modes and the choice experiment attributes. Important to note, studies emphasizing automated vehicles (where the ride may be driverless) are not included in this analysis as the perceptions of sharing attributes can be highly affected by the automation feature (e.g. Krueger et al., 2016, Etzioni et al., 2021). Additionally, we exclude portfolio-based Mobility as a Service studies where microtransit (and related attributes) is a minor focus (e.g. Caiati et al. (2020)).

**Table *1*** shows that each study covers different definitions of microtransit-like services, with different sets of attributes and information provided to customers: Yan et al. (2018) provide survey-takers information about additional pickups; Frei et al. (2017) include headway for their flexible route, demand-responsive transit; Chavis and Gayah (2017) feature the availability of GPS tracking of vehicle for more traveler information; Al-Ayyash et al. (2016) consider in-vehicle WiFi capabilities. The two remaining studies, Alonso-González et al. (2020b) and Alonso-González et al. (2020a), were derived from the same survey. In Alonso-González et al. (2020b), the choice experiment included uncertainty for the waiting and in-vehicle times to determine Values of Time for individual and shared rides. Alonso-González et al. (2020a) then considered mode choice of flexible transit alternatives specifically. On the whole, the most common attributes included in choice experiments regarding microtransit are the out-of-vehicle travel time and additional passengers sharing the ride.

The current study is unique in including attributes for a minimum reservation time and availability of a sheltered boarding location akin to a bus shelter. By knowing the demand for trips with an earlier notice, service providers can better optimize vehicle routing and possibly pool more trips together. Since curb-to-curb services rely on travelers walking to a boarding location, we hypothesize that shelter availability from adverse weather may significantly impact the utility of our pooled ride alternatives.

**Table 1 Stated Choice Studies of Microtransit Options and Attributes**

| Author(s) | Country | Alternatives (dep. var) | Cost | IVTT | OVTT | Time Uncertainty | Headway | Additional Pax | Transfers | Additional Pickups | Integrated Advanced Technology | Minimum Reservation Time | Sheltered Boarding Location |
|---|---|---|---|---|---|---|---|---|---|---|---|---|---|
| Yan et al. (2018) (N=1,163) | USA | Drive, Mtransit, Bike, Walk (choice) | | ✓ | ✓ | | | | ✓ | ✓ | | | |
| Frei et al. (2017) (N=183) | USA | Car, Transit, Flexible transit (choice) | ✓ | ✓ | ✓ | | ✓ | | ✓ | | | | |
| Chavis and Gayah (2017) (N=173) | USA | Fixed route transit, flexible transit, solo driving (choice) | ✓ | ✓ | ✓ | | | | | | ✓ | | |
| Al-Ayyash et al. (2016) (N=1,393) | Lebanon | shared-ride taxi (weekly frequency) | ✓ | ✓ | ✓ | | | | ✓ | | ✓ | | |
| Alonso-Gonzalez (2020a) (N=1,006) | Netherlands | Individual and Pooled (choice) | ✓ | ✓ | | | | | ✓ | | | | |
| Alonso-Gonzalez (2020b) (N=1,006) | Netherlands | Combined modes: flexi, flexi + bus, bus + bus (choice) | ✓ | ✓ | ✓ | ✓ | | | ✓ | | | | |
| Current Study (N=1,326) | Israel | Current mode (car vs. Transit), MT-S, MT-V (choice) | ✓ | ✓ | ✓ | | | ✓ | | | | ✓ | ✓ |

## 2.4 COVID-19 Effects on Shared Transportation

The Coronavirus pandemic has tremendously impacted travel via both supply and demand effects. Because the risk of exposure is a function of physical proximity, many countries enacted large-scale lockdowns, limited access to public spaces, and imposed social distancing directives. These lockdowns have dramatically decreased the demand for travel (Glanz et al., 2020). While demand for mobility has grown since initial lockdowns, public transit ridership has yet to recover (Rothengatter et al., 2021). Public transit was significantly impacted, with some agencies reporting a 90% decrease in ridership (Verma, 2020, De Vos, 2020, Abdullah et al., 2020). Connected to this, attitudes towards public transit relating to COVID-19 are revealed to be negative (Thomas et al., 2021). Ridehailing demand was similarly impacted, with an 80% decrease in ridership (Higgins and Olson, 2020). While ridehailing remained operational during the pandemic for essential travel, one of the first actions of TNCs was to halt ridepooling operations (e.g. UberPool and Lyft Line) (Bond, 2020).

As the pandemic evolved and lockdowns gradually eased, travel behavior is still impacted by the virus-related risk perceptions and contraction risk. Travelers will continue to evaluate the tradeoffs between the need to travel (e.g., to maintain livelihoods) and being exposed to COVID-19 in shared rides (Borowski et al., 2021, Rahimi et al., 2021). Ongoing work is examining the evolving perceptions and priorities of travelers in the uncertain COVID-19 era. Said et al. (2021) indicate there has been a change in intention to use pooled modes due to the pandemic. Another recent study found that approximately 41% of survey-takers would consider using ridehailing even if operators take extra precautions by providing masks, gloves, and sanitizing gel, whereas only 28% would be willing to pay more for the added protective measures (Awad-Núñez et al., 2021). The percentage of those willing to use public transit under the same conditions was similar. A Toronto survey found that 15% of respondents declared an intention never to use ridehailing again, and 21% would never use ridepooling (Loa et al., 2020). From the same report, approximately 30% of riders prefer to wait until the virus is no longer a threat as the earliest point in time when they would consider using ridehailing or pooling. In general, travelers are moving from public to private modes (Das et al., 2021). Indeed, several negative long-term consequences include the persistent reluctance to use shared modes, rebounding of car travel, and increase in car purchases (Hensher, 2020). On the other hand, the role of risk-perceptions and user intentions over time and across cultures is less well understood.

## 2.5 Summary and Literature Take-aways

The pandemic has shown that ridership habits can shift rapidly and may rebound as circumstances change. A Chicago study found that 80% of lapsed riders intend to return to transit (RTA, 2021). We noted a limited number of papers on microtransit adoption and limited overlap with pandemic travel analysis. Specifically, it is still unclear how users view travel in shared vehicles and where microtransit fits with evolving risk perceptions. The proposed research will help map out identity and values surrounding sharing and risks surrounding pandemic travel. Specifically, the paper analyzes the willingness to engage in microtransit and how different serviced features such as vehicle sizes and seating configurations related to the ability to socially distance, affect the demand for different service models.

# 3 Data

The data were collected using a SC survey in Tel Aviv, Israel. The survey was distributed to car and transit commuters throughout the metropolitan region which comprises nearly half of Israel's 9M population and includes the core city of Tel Aviv, the main business, culture, and high-tech hub. Tel Aviv also operates a small-scale microtransit pilot service known as Bubble-Dan, which operated before and during the pandemic (Bubble-Dan, 2021). A screening was applied to include only participants who commute at least three times a week with a commute duration of at least 10 minutes using only a personal vehicle or public transit. Data about current commute attributes, socio-demographics, past and expected future life events, latent attitudes, and choice experiments with microtransit alternatives were collected. Using the respondents' current commute attributes, we determine their commute mode, cost, and travel time for the reference alternative in the CE—referred to as the "Status Quo" (SQ).

Latent attitudes were measured using survey item statements based on respondent's sharing experience, schedule-keeping, environmental stances, and comfort with situations related to risks of COVID transmission. Sharing attitudes and COVID Comfort statements used in this modeling are summarized in **Table** *2*. The sharing-related items are drawn from previous research and modified to orient them around the sharing economy. The COVID-19 related items were created specifically for this survey. Each item uses a 5-point Likert scale that ranges from "Strongly Disagree" to "Strongly Agree" (Lehmann and Hulbert, 1972). In addition, we asked respondents to report the degree to which the COVID-19 pandemic had affected their lives. Respondents were asked to respond to this question by indicating "No Change, Little Change, Not Sure, "Big Change," or "Very Big Change." These questions were then used to identify latent variable effects on microtransit decision-making.

The experiment design and implementation was developed in sequential steps (**Figure** *2*) following best practice guidance (Johnson et al., 2013, Kløjgaard et al., 2012). Step 1 covered qualitative attribute development, focusing on identification, selection, and presentation. Following literature and industry report analysis, seven attributes were selected, representing two microtransit vehicle sizes. Further informal testing in step 2 led to a fractional factorial experimental design with six microtransit scenarios (see more details in Soria et al. (2019)). Given Israel's limited familiarity with microtransit services, several auxiliary questions were designed to measure attribute acceptance cutoffs, importance, and choice certainty. In step 3, a full web survey implemented in Qualtrics was administered to 301 pilot respondents.

**Table 2 Sharing and COVID-19 Comfort Items with Coding**

| Item | Coding | Source |
| --- | --- | --- |
| I enjoy using sharing economy services | SI1_enjoy | Van der Heijden (2004) |
| I can see myself increasing my use of shared mobility in the future | SI2_increase | Bhattacherjee (2001) |
| I have never had a bad experience using sharing economy services | SI3_exp | Current study |
| Inclusion of Other in Self (IOS) | IOS | Adapted from Aron et al., (1992) |
| Given the current situation caused by the COVID-19 outbreak, I would feel comfortable engaging in the following activities: | - | Current study |
|    Ridesharing with strangers | CC1_ride | |
|    Eating out at a restaurant. | CC2_rest | |
|    Going to the grocery store | CC3_grocery | |

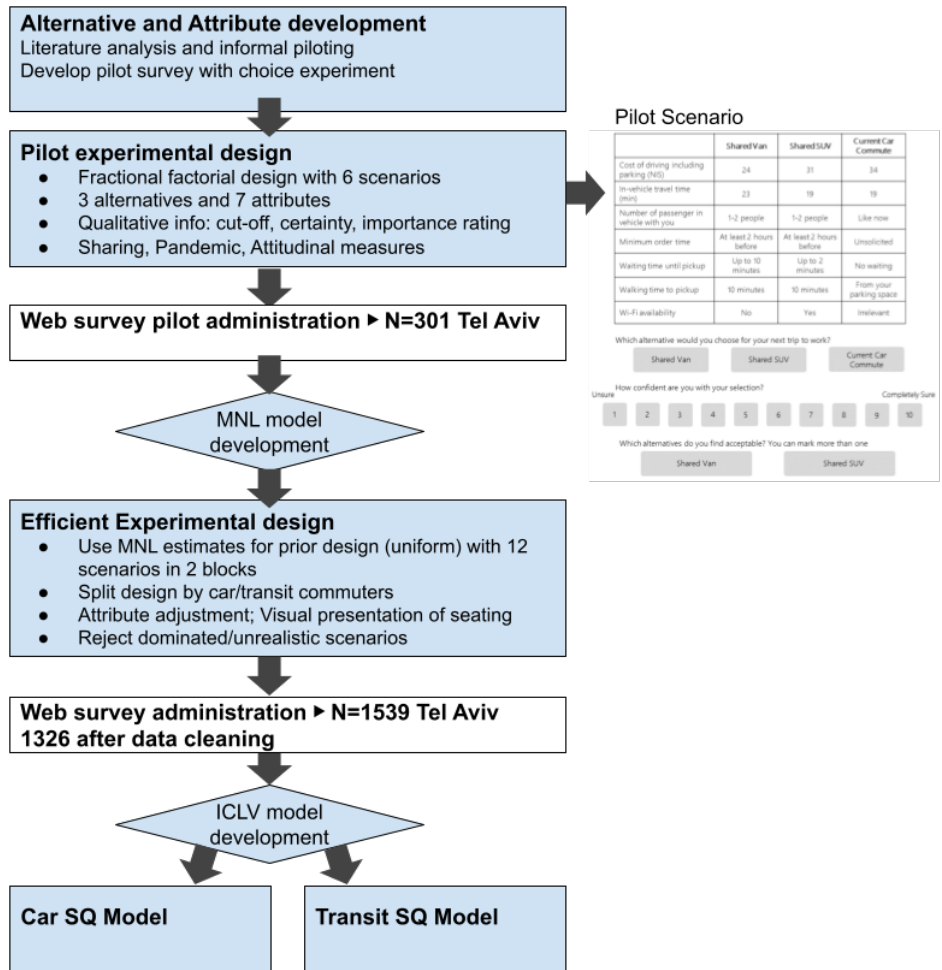

**Figure 2 Steps in Development of a Discrete Choice Experiment**

Results were analyzed in step 4 using discrete choice modeling, leading to the development of priors for an efficient experimental design in step 5 using Ngene software (ChoiceMetrics 2012). The resulting design included three alternatives: Status Quo (SQ, either car or public transit), Microtransit Sedan (MT-S), and Microtransit Van (MT-V). The pilot analysis led to broadening attribute ranges and providing a visual presentation for the seating variable. The travel time and travel cost attributes for current travel alternatives were pivoted off the reported (RP) levels to improve the realism of the experiment (Hensher and Rose, 2007, Train and Wilson, 2008, Etzioni et al., 2020). **Table *3*** lists the mode attributes included in the final experiment along with the attribute levels. Graphics were presented in the choice experiment to reflect the number of additional passengers and which seats are available (**Figure *3***). The graphical presentation of seating designation and vehicle seating configuration allows more direct understanding of precise links to mode-pooling decisions (Etzioni et al., 2021). The respondents' current travel cost and time were defined using the following logic. If the typical commute mode is driving, the respondent provides further information about parking such as search time, if there is a reserved parking area, and if they pay for that parking. The travel costs are approximated for drivers by summing the daily parking fee and their travel distance in kilometers multiplied by two, using this information. In Israel, the value of 2 ILS per km is a gross estimate used by the public sector for reimbursing direct car use expenditures and is also

**Table 3 Choice Experiment Alternative Attribute Levels**

|  | Status Quo (fixed) | Microtransit Sedan | Microtransit Van |
|---|---|---|---|
| Cost (per day) | Current Cost | -10%/-20%/-30% (CAR) | -15%/-30%/-45% (CAR) |
|  |  | +75%/+125%/+175% (PT) | +50%/+100%/+150% (PT) |
| Travel time | Current Door-to-door Time | -30%/ 0 / +30% (CAR) | 0/+15%/+30% (CAR) |
|  |  | -30%/ 0 / +30% (PT) | 0/-15%/-30% (PT) |
| Number of occupants in a vehicle |  | 1 person (driver)/ 2 people/ 4 people | 1 person (driver)/ 5 people/ 8 people |
| Minimum Reservation Time Before Boarding |  | 2hr/10 min/5 min before | 2hr/10 min/5 min before |
| Waiting Time |  | 2 min/up to 5 min/up to 10 min | 2 min/up to 5 min/up to 10 min |
| Walking Time |  | No walking/Up to 5 min walk/Up to 10 min walk | No walking/Up to 5 min walk/Up to 10 min walk |
| Station amenity |  | Designated-shelter (yes/no) | Designated-shelter (yes/no) |

**Microtransit Sedan**

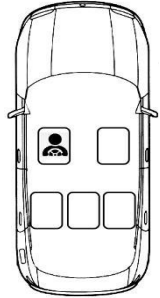 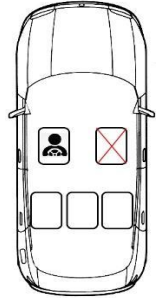 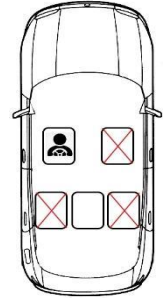

1 person (driver)          2 People          4 People

**Microtransit Van**

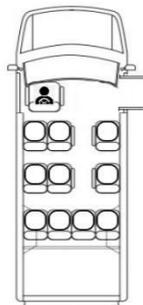 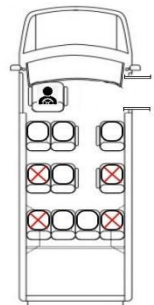 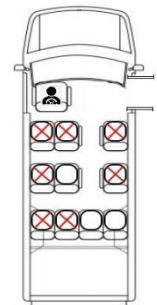

1 person (driver)          5 people          8 people the maximal value allowed by the Ministry of Transport regulations for determining direct cost-sharing in voluntary carpooling arrangements between driver and passengers. Travel time for car commuters is the sum of their stated commute time and parking search time. Travel cost corresponds to the single trip fare for transit commuters, and travel time is their current stated commute time. Because both car and transit commuters responded to this survey, the design was optimized for each commuter group separately.

**Figure 3 Depiction of Additional Passengers in Choice Experiment**

The experiment is based on a D-efficient Bayesian design created using Ngene (ChoiceMetrics, 2012, Yu et al., 2011). The a priori coefficient values (Rose et al., 2008) were obtained using uniform distributions from the pilot survey (Soria et al., 2019). However, this pilot survey considered only car commuters; hence we assumed that the coefficient values for all attributes were equal across groups. Dominated and unrealistic alternatives were excluded using the Federov algorithm (ChoiceMetrics, 2012). These actions are put in place to ensure designs are plausible and realistic. For example, given that the transit fare was relatively low, the fare for the microtransit alternatives was constrained to be greater than the transit one for transit commuters. In contrast, with car commuters, costs for MT-S and MT-V were always lower than the car cost. The design extracted 12 choice scenarios for each SQ mode; however, the scenarios were randomly assigned to two fixed sets of 6 scenarios to prevent respondent fatigue (Caussade et al., 2005).

In step 6, a web-based respondent panel was used to collect 1539 survey responses in May 2020. The data were cleaned by first removing responses that did not complete the choice experiment portion. To preserve data quality, responses that took less than 5 minutes or showed patterns of inattentiveness were removed. Because the average time it took to complete the survey was approximately 30 minutes, we treated 5 minutes as the minimum cutoff to complete it earnestly. We further carried out qualitative pattern analysis and removed non-differentiated ratings in blocks of questions (straight lining) (e.g. Yan 2008). For the current analysis, responses with current commute times greater than 90 minutes were excluded to decrease heterogeneity and maintain a reasonable commuter service area for microtransit. After cleaning and subsetting the database, 1326 responses (86%) were retained, resulting in 7956 choice experiment observations. Of these 1326 responses, there were 879 (66%) car and 447 (34%) transit commuters.

**Table 4** provides the descriptive statistics for the observed variables, COVID-19 impacts, and attitudes. For the respondents' current commutes, the largest difference between groups is the travel cost. Car commuters are also more likely to be male, married and have more children. The voter variable is a dummy variable denoting if the respondent voted in the 2020 legislative elections in Israel, for which there is little difference between groups. The Inclusion of Other in the Self (IOS) scale measures how close the respondent feels with strangers (Aron et al., 1992). In this study, we specifically asked respondents how close they feel to a stranger sharing a pooled vehicle. Overlapping circles are used to depict IOS, where more overlapping circles denote closer connection with other riders (**Figure 4**). Car commuters, somewhat surprisingly, rates higher on this scale. The pandemic has similarly impacted both groups. Lastly, they share nearly the same attitudes towards the sharing economy and COVID Comfort, with the most significant difference being the CC1_ride. Transit commuters are more comfortable sharing a ride with a stranger during the pandemic.

**Table 4 Descriptive Statistics of Modeling Variables**

| Variable | All Commuters (std. deviation) | Car Commuters (std. deviation) | Transit Commuters (std. deviation) |
|---|---|---|---|
| Current Commute | | | |
|   Travel Cost (ILS) | 35.31 (38.98) | 50.34 (40.23) | 5.78 (2.94) |
|   Travel Time (minutes) | 33.20 (15.75) | 31.59 (14.71) | 36.37 (17.18) |
| Individual Descriptors | | | |
|   Married | 55.81% | 61.43% | 44.74% |
|   Gender is Male | 50.45% | 53.12% | 45.19% |
|   Voter | 89.97% | 89.30% | 91.28% |
|   Number of Children | 1.27 (1.63) | 1.45 (1.60) | 0.90 (1.64) |
| COVID Impact | | | |
|   No Impact | 1.96% | 1.37% | 3.14% |
|   Little Impact | 32.81% | 31.63% | 35.12% |
|   Big Impact | 41.86% | 42.54% | 40.49% |
|   Very Big Impact | 12.67% | 12.97% | 12.08% |
|   Not Sure | 10.70% | 11.49% | 9.17% |
| Attitudes | | | |
|   IOS (min = 1, max = 7) | 2.75 (1.60) | 2.80 (1.64) | 2.67 (1.52) |
|   SI1_enjoy | 3.05 (1.08) | 2.97 (1.08) | 3.20 (1.06) |
|   SI2_increase | 3.17 (1.04) | 3.12 (1.06) | 3.27 (1.01) |
|   SI3_exp | 3.25 (1.04) | 3.25 (1.04) | 3.26 (1.05) |
|   CC1_ride | 2.27 (1.11) | 2.10 (1.04) | 2.59 (1.16) |
|   CC2_rest | 2.46 (0.83) | 2.45 (0.81) | 2.49 (0.85) |
|   CC3_grocery | 3.66 (0.96) | 3.63 (0.96) | 3.75 (0.95) |
| N | 1326 | 879 | 447 |

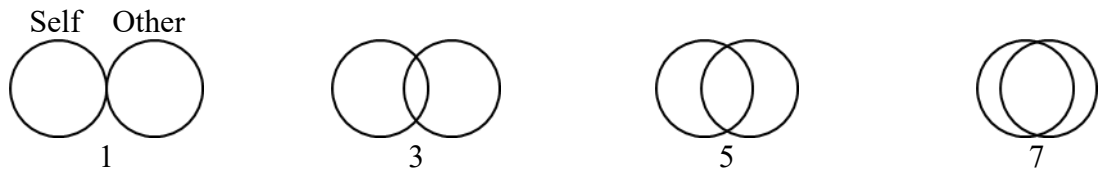

**Figure 4 Inclusion of Other in the Self (IOS) Scale**

# 4 Methodology

The purpose of this research was to identify the acceptability and tradeoffs among novel microtransit attributes and quantify the effect of latent variables on the decision-making process. Separate models were estimated using an Integrated Choice and Latent Variable (ICLV) model for the two commuter groups. The ICLV framework allows the choice and latent variable models to be estimated simultaneously (Temme et al., 2008, Bolduc and Alvarez-Daziano, 2010, Abou-Zeid and Ben-Akiva, 2014). **Figure 5** depicts the theorized relationship between the latent variables, mode attributes, utility of each mode, and, finally, mode choice. To estimate the ICLVs, we follow the guidelines from Walker (2001). From the guidelines, the first steps are to identify the choice model and structural equation model separately. Once this is completed, the models are jointly estimated.

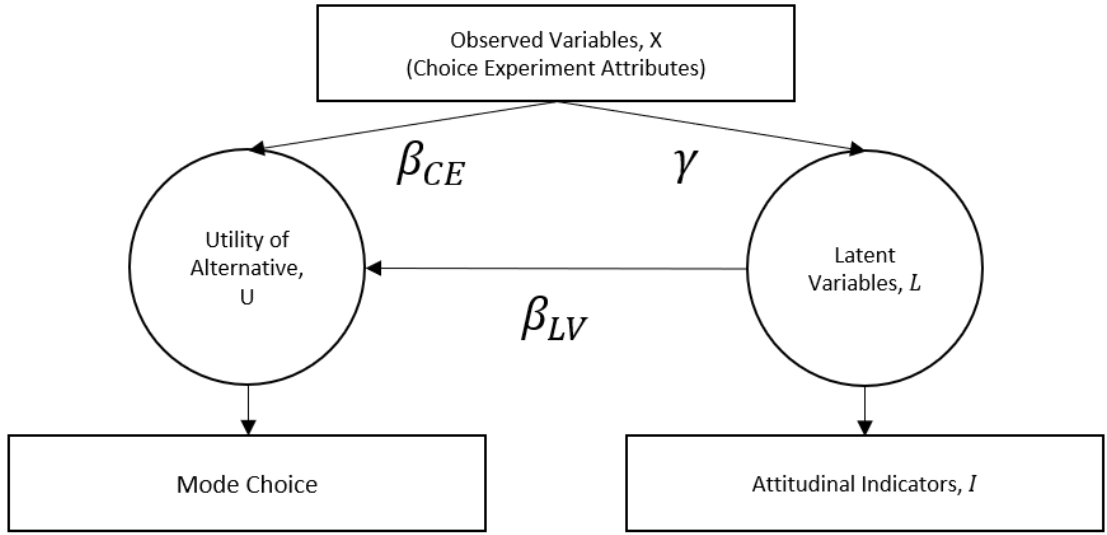
**Figure 5 Integrated Choice and Latent Variable Framework**

## 4.1 Discrete Choice Model

The first step of the guideline is to identify the utility specification of the choice model correctly. We completed this step by estimating a Multinomial Logistic Regression (MNL) for each commuter group with PandasBiogeme (Bierlaire, 2018). Equations 1a and 1b describe the general utility specification. $U_{in}$ is the latent utility of alternative I of observation n, $X_{CE}$ is the matrix of explanatory variables from the choice experiment, L are the latent variables, $\beta_{CE}$ and $\beta_{LV}$ are the corresponding coefficients, and $\epsilon$ is the independently and identically distributed (IID) error term.

$$U_{in} = V_{in}(X, L; \beta) + \epsilon_{in} \qquad (1a)$$

$$U_{in} = \beta_{CE} X_{CE} + \beta_{LV} L + \epsilon_{in} \qquad (1b)$$

## 4.2 Structural Equational Model for Attitudinal Indicators

After identifying the mode choice model, the second step is to identify the latent variables. We estimated a Structural Equation Model (SEM) using the attitudinal items in the measurement component and explanatory variables including, socio-demographics, experience with sharing economy services, and the structural component's life impact and comfort related to COVID-19. Equations 2a and 2b describe the measurement and structural components, respectively. The SEM's were first estimated using the R package *psych*, then confirmed again using PandasBiogeme (Revelle, 2018, Bierlaire, 2018). With both choice and latent variable models identified, the last step is to estimate the integrated models simultaneously.

Equations 2a (structural) and 2b (measurement) below describe the SEM. L is the latent variable, the intercept $\theta$, observed variables $X_{LV}$, the corresponding estimated coefficients $\gamma$, and the error term, $\eta$, which is IID multivariate normally distributed. I is the response for the attitudinal items listed in **Table 2**. It is a function of $\alpha$ an intercept, $\lambda$ the estimated coefficients, L a matrix of latent variables estimated from Equation 2, and $\zeta$ the IID multivariate normal error term. $\sigma$ is a random variable to capture the random taste heterogeneity of the sample and is added to estimate numerically the likelihood described in the following subsection.

Several latent variables were estimated representing the respondents' attitudes towards Environmental Sustainability, Schedule Making, Pro-Sharing Economy, and COVID Comfort.

Only the last two were consistently significant in at least one commuter group, with a hierarchical relationship shown in **Table *2***. The latent variables were validated with the following metrics and threshold values: Comparative Fit Index (CFI) > 0.90, Root Mean Square Error of Approximation (RMSEA) < 0.06 and Standardized Root Mean Square Residual (SRMR) < 0.08 following recommendation in literature (Hu and Bentler, 1999, Hooper et al., 2007).

$$L = \theta + X_{LV}\gamma + \eta \tag{2a}$$

$$I = \alpha + \lambda L + \sigma + \zeta \tag{2b}$$

### 4.3 Integrated Choice and Latent Variable Model

The models are estimated simultaneously by maximizing the joint log-likelihood of each component. Equation 3 shows the joint likelihood. This integrand cannot be solved analytically, so it was estimated numerically with random variables, $\phi$, in the latent variable model. $p(X, L; \beta)$ is the likelihood from the standard MNL. $f(L, X_{LV}; \gamma)$ is the likelihood from the structural component of the SEM and $g(I, L, \sigma; \lambda)$ is the likelihood of the measurement component.

$$Likelihood = \prod_{n=1}^{N} \int_{L} p(X, L; \beta) f(L, X_{LV}; \gamma) g(I, L, \phi; \lambda) dL \tag{3}$$

## 5 Results

Two ICLV models were estimated, one for car commuters and one for transit commuters, and the results are shown in **Table *5*** and **Table *6***. For added clarity, the structure of the latent variables in the ICLVs is shown in **Figure *6***. Following extensive specification testing done individually, the models were similarly specified so that the results were as directly comparable as possible. Mode attributes were limited to the discrete choice portion of the ICLVs while attitudinal items and sociodemographic variables were limited to the latent variable models. Additionally, the latent variables were hypothesized to exist for both commuter groups and, subsequently, share the same scales. The final utility specifications are described in Equations 4a to 4d. Equation 4a shows COVID Comfort in the utility specification for the car alternative. Two latent variables were identified and included in the final model because the Pro-Sharing Economy construct was found to indirectly affect the utility of car commuters via a structural relationship with COVID Comfort as shown in ***Figure 6***.

$$V_{car} = \beta_{Car} + \beta_{Car,Cost}CarCost + \beta_{Car,Time}CarTime \tag{4a}$$
$$+ \beta_{CovidComfort}CovidComfort$$

$$V_{PT} = \beta_{PT} + \beta_{PT,Cost}PTCost + \beta_{PT,Time}PTTime + \beta_{CovidComfort}CovidComfort \tag{4b}$$

$$V_{MTS} = \beta_{MTS} + \beta_{MTS,Cost}MTSCost + \beta_{MTS,Time}MTSTime + \beta_{MTS,Walk}MTSWalkTime \tag{4c}$$
$$+ \beta_{MTS,Wait}MTSWaitTime + \beta_{MTS,MinRes}MTSMinResTime$$
$$+ \beta_{MTS,Passengers}MTSPassengers + \beta_{MTS,Shelter}MTSShelter$$

$$V_{MTV} = \beta_{MTV} + \beta_{MTV,Cost}MTVCost + \beta_{TV,Time}MTVTime \tag{4d}$$
$$+ \beta_{MTV,Walk}MTVWalkTime + \beta_{MTV,Wait}MTVWaitTime$$
$$+ \beta_{TV,MinRes}MTVMinResTime + \beta_{MTV,Passengers}MTVPassengers$$
$$+ \beta_{MTV,Shelter}MTVShelter$$

**Table 5 Microtransit Choice Models Results**

| | Alternative | | | | | |
|---|---|---|---|---|---|---|
| | Car | | | Public Transit | | |
| Coefficient | Car (Std. Error) | MT-S (Std. Error) | MT-V (Std. Error) | Transit (Std. Error) | MT-S (Std. Error) | MT-V (Std. Error) |
| Constant | 0 - fixed | -4.31** (0.255) | -3.58** (0.272) | 0 - fixed | -1.56** (0.263) | -1.67** (0.267) |
| Travel cost (ILS) | -0.00319** (0.000954) | [1]-0.0068** (0.00207) | [1]-0.00546* (0.00277) | -0.116** (-0.0341) | -0.0423** (0.0157) | -0.0683** (0.0174) |
| In-vehicle travel time (minutes) | -0.0494** (0.00509) | -0.0312** (0.00455) | -0.0364** (0.00444) | -0.0386** (0.00592) | -0.029** (0.00614) | -0.0248** (0.00675) |
| Walk time (minutes) | - | -0.0450** (0.0111) | -0.110** (0.0143) | - | NS | NS |
| Wait time (minutes) | - | -0.0306* (0.0143) | -0.0691** (0.0176) | - | NS | NS |
| Minimum reservation time before boarding (minutes) | - | -0.00140* (0.000680) | -0.00363** (0.00105) | - | -0.00348** (0.00131) | -0.0064** (0.00116) |
| Number of people in vehicle | - | NS | -0.0872** (0.0276) | - | -0.0978** (0.0381) | -0.0915** (0.0315) |
| Sheltered Boarding Location | - | NS | NS | - | NS | 0.378** (0.104) |
| COVID Comfort | -1.34** (0.0969) | - | - | -0.14^ (0.0828) | - | - |
| n observations | 5274 | | | 2682 | | |
| $\rho^2$ | 0.164 | | | 0.296 | | |
| Final Loglikelihood | -47665.05 | | | -13590.62 | | |

(NS) Not statistically significant at $\alpha$ = 0.1, not estimated in final model
(^) significant at $\alpha$ = 0.1
(*) significant at $\alpha$ = 0.05
(**) significant at $\alpha$ = 0.01
[1] Interacted with dummy variable for having commute time greater than 65 minutes, otherwise statistically insignificant

**Table 6 Latent Variable Models Results**

| Coefficient | Model | |
|---|---|---|
| | Car (Std. Error) | Transit (Std. Error) |
| **COVID Comfort** | - | - |
| CC1_ride | 1 – fixed | 1 – fixed |
| $\alpha_{CC1}$ | - | - |
| CC2_rest | 0.496** (0.0323) | 0.626** (0.0401) |
| $\alpha_{CC2}$ | 1.41** (0.0696) | 0.865** (0.106) |
| CC3_grocery | 0.643** (0.0355) | 0.649** (0.0425) |
| $\alpha_{CC3}$ | 2.27** (0.0763) | 2.07** (0.112) |
| $\theta_{CC}$ | 1.02** (0.122) | 2.46** (0.111) |
| Impact - Unsure | -0.572** (0.0976) | 0.0887 (0.123) |
| Impact - No Change | 0 – fixed | 0 – fixed |
| Impact - Little Change | -0.539** (0.0939) | 0.333** (0.111) |
| Impact - Big Change | -0.916** (0.0957) | 0.0223 (0.112) |
| Impact - Very Big Change | -1.32** (0.102) | -0.383** (0.122) |
| Married | -0.138** (0.0220) | - |
| Male | - | 0.186** (0.0389) |
| Number of Children | - | -0.0472** (0.0116) |
| $\sigma_{CC}$ | 0.113^ (0.0641) | 0.628** (0.0259) |
| **Pro-Sharing Economy** | 0.661** (0.0302) | - |
| SI1_enjoy | 1 – fixed | - |
| $\alpha_{SI1}$ | - | - |
| SI2_increase | 0.932** (0.0310) | - |
| $\alpha_{SI2}$ | -0.355** (0.0939) | - |
| SI3_exp | 0.592** (0.0283) | - |
| $\alpha_{SI3}$ | 1.49** (0.0856) | - |
| $\theta_{SI}$ | 2.21** (0.0469) | - |
| Ridehailing App Experience | 0.312** (0.0252) | - |
| Carpooling App Experience | 0.292** (0.0249) | - |
| Carsharing App Experience | 0.0301** (0.00621) | - |
| IOS | 0.110** (0.00731) | - |
| Voter | 0.201** (0.0371) | - |
| Male | 0.112** (0.0230) | - |
| $\sigma_{SI}$ | 0.609** (0.0159) | - |
| $\chi^2/df$ | 2.67 | 2.47 |
| CFI | 0.923 | 0.939 |
| RMSEA | 0.044 | 0.055 |
| SRMR | 0.032 | 0.028 |

(-) Not applicable
(^) significant at $\alpha = 0.1$
(*) significant at $\alpha = 0.05$
(**) significant at $\alpha = 0.01$

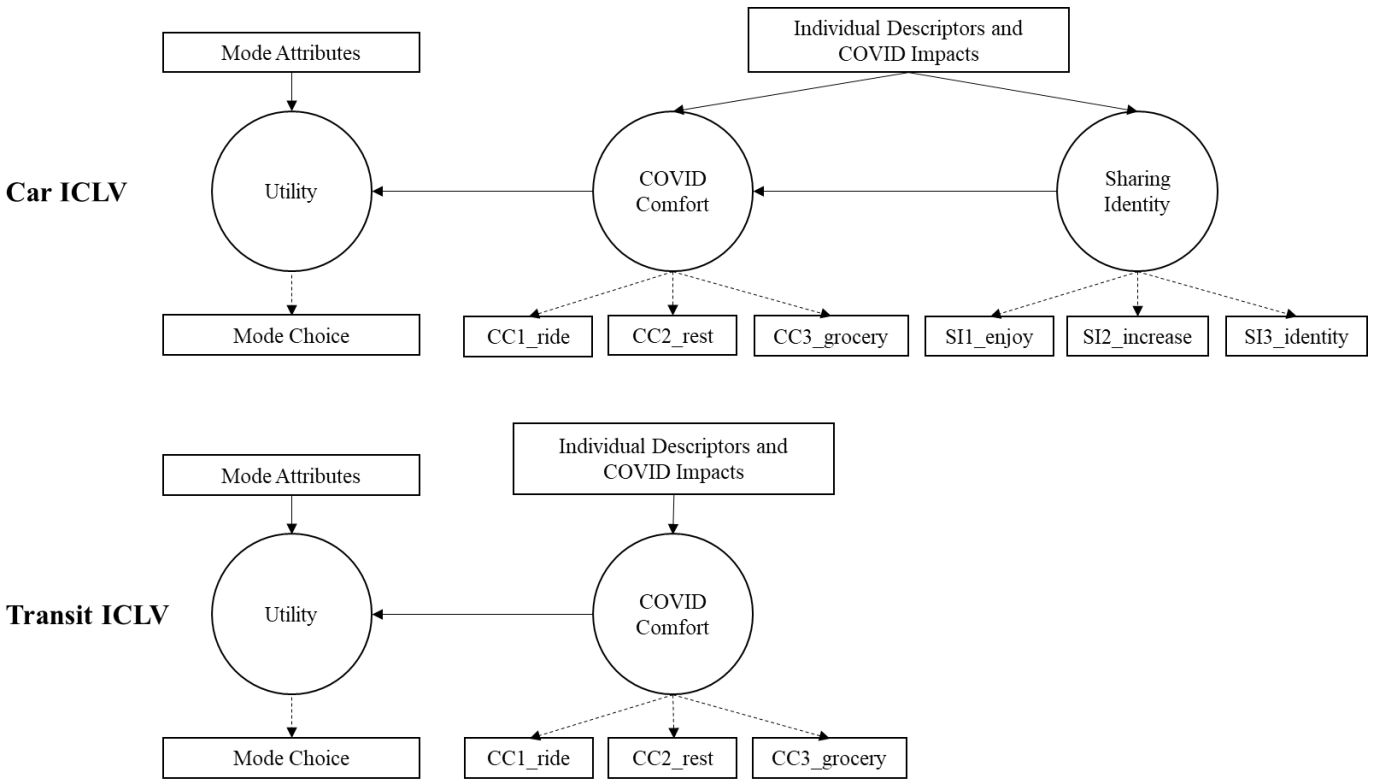

**Figure 6 Structure of ICLV Models for Each Commuter Group**

## 5.1 Microtransit Acceptance and Mode Attributes

All mode attributes for all alternatives were included in the discrete choice portion of the ICLV during model development. In the final estimations, statistically insignificant variables were not included. **Table 5** intercepts show an inherent attraction to the status quo modes. It was expected that transit commuters would find the on-demand sharing alternatives inherently more attractive than the status quo; however, the results show the opposite. When examining the attribute effects, all features have the expected sign, while we note several differences between the two commuter groups. The more traditional travel cost and travel time attributes are statistically significant for the status quo choice. Yet, we note that transit commuters have a higher sensitivity to cost and lower sensitivity to time than solo drivers, which is true also for microtransit options, suggesting that transit commuters transfer their preferences onto new options. Generic travel cost sensitivities for drivers were not significant, prompting us to test interactions with other variables to examine segmentation. Unlike other work suggesting marginally decreasing sensitivities (Daly, 2010), we find that only drivers with long commutes (greater than 65 minutes) are consistently sensitive to the cost attribute, with a higher sensitivity for the MT-S and MT-V. We speculate that the limited cost sensitivity is due to drivers' difficulty to perceive the largely hidden cost of driving and parking, coupled with an experimental design effect where prices for the microtransit alternatives were set to be lower than solo driving (Andor et al., 2020, Shoup, 2021).

The novel attributes were only included for the microtransit alternatives, and on the whole, we note that car commuters are sensitive to a greater range of microtransit attributes than transit commuters. This likely reflects the fundamental dissimilarity between driving and microtransit, which comports several unfamiliar attributes (Alemi et al., 2018). Specifically, we note a major difference for time-related attributes of walking time to the curbside pickup location, and waiting time until pickup. Car commuters appear to have a strong aversion to walking and waiting, with a strong penalty for the van option. Instead, transit users are only sensitive to the in-vehicle travel time, with insignificant walking and waiting parameters, similar to the results from Frei et al. (2017). The last time-related attribute, minimum reservation time, is statistically significant for both alternatives, albeit with a lower magnitude than other time measures. Instead, a few features come into play only for the transit commuters. The number of additional passengers matters for both vehicle sizes in the transit model while only impacting the larger MT-V among drivers. Finally, the 'sheltered boarding' attribute is only significant for MT-V in the transit model. Additionally, car commuters are not sensitive to the number of additional passengers for MT-S, possibly stemming from this mode being a familiarly sized vehicle with relatively low capacity. Alonso-González et al. (2020a) also posit that perceptions of sharing reach a tipping point at four additional passengers since a vehicle larger than a regular car is needed. Finally, the presence of a sheltered boarding location is only statistically significant for MT-V.

In summary, it is likely that transit users have interiorized the *transit-like* attributes of walking and waiting that are intrinsic to scheduled services. This is reflected in the lack of significant effects for transit commuters. Instead, we note that transit users appear more sensitive than drivers to reservation time, shelter, and the number of other passengers, attributes that are more affected by the ICT-supported mobility platform and the smaller vehicle sizes than experienced in current transit travel.

### 5.2 Latent Variable Effect

Overall, model fit for both latent variable models indicates good fit with CFI > 0.90, RMSEA < 0.06 and SRMR < 0.08 in both models (Hu and Bentler, 1999, Hooper et al., 2007). The most evident difference between car and transit commuters is in the latent variable portion of the ICLVs. The structure of the latent variables in the ICLV models is illustrated in **Figure 6**. As shown by estimates in **Table 6**, decision-making by car commuters was affected by both latent variables, namely: *Pro-Sharing Economy* and *COVID Comfort*, while transit users, surprisingly, were not motivated by these factors. COVID Comfort is designed to represent a respondent's comfort with different COVID-19 risk situations. As such, respondents' comfort in grocery stores, eating in restaurants, and sharing a vehicle with a stranger are used to identify this latent variable. The COVID Comfort parameter in **Table 5** shows a negative effect. That is, the more at ease respondents are with these situations, the more likely they are to accept trying the microtransit services. The structural component of this latent variable consists mostly of variables reflecting the impact the pandemic has had on respondents' lives, measured by the *Impact* variable. The impacts are ordinal in nature; however, here, we chose to model the impact as a discrete categorical variable to facilitate separate modeling of the opt-out where respondents indicate their uncertainty. This decision proved to be useful as those who were uncertain of COVID impacts were found to be less comfortable with COVID than those who experienced "Little Change." Additionally, it was advantageous because the jump in effect from "No Change" to "Little Change" in the car commuter group resulted in a larger impact than the jump from other levels. We also considered the risk of transmission to significant others related to a respondent and found that married people are less comfortable with risky COVID situations. Transit commuters who had "Little Change" in their lives from COVID were more comfortable with it than those who had had impacts at other levels or were unsure about its impact. We also considered the risk of COVID transmission to loved ones and found that families with more children were less likely to be comfortable with COVID exposure situations. Lastly, we found that men tended to be more comfortable with COVID-19 risk situations, which resonates with observations that men are less concerned about virus contraction and less likely to get vaccinated (Galasso et al., 2020, Lazarus et al., 2021)

Modeling also reveals that COVID Comfort is directly affected by the Pro-Sharing Economy construct (albeit only for drivers, as depicted in **Figure** *6*). Because sharing in this context is of physical assets (including public areas), we hypothesized a structural relationship between these two latent variables. The positive sign implies that experience with sharing economy services — used to measure higher Pro-Sharing Economy — is underpinning higher comfort with sharing resources during COVID-19. There are two issues to note here. First, the hypothesized hierarchical causation suggests that sharing is an established trait that affects how respondents behave in the novel and temporary context of pandemic social distancing. In practice, it is likely that the evolving objective and subjective risks, as well as experience and fatigue from social distancing, will continue to shape willingness to ridepool. Second, we expected Pro-Sharing Economy to be a driving factor for transit users. Instead, we could find no evidence of this construct affecting neither COVID Comfort nor likelihood to use microtransit directly. We speculate that the transit users we observe, especially during COVID-19, are not choice-riders driven by sharing ideals but rather motivated by practical necessity. Like above, there are likely to be dynamic effects at play, connecting ridership to changing employment circumstances and COVID-19 risk levels. These issues warrant further research.

In addition to the sharing economy constructs, the IOS scale is used to measure sharing propensity. Our study finds that the more closely a respondent identifies with other riders, the higher they score on Pro-Sharing Economy. Several personal characteristics are found to be related to sharing ideals. Being a voter in the latest election is positively correlated with sharing. We speculate that voters may have higher civic duty orientation associated with higher sharing identities (Fowler, 2006, Bolsen et al., 2014). Lastly, men tend to have higher sharing identities, and we attribute this to women's perceptions of (lack of) safety, especially in situations where personal space cannot be guaranteed (Morales Sarriera et al., 2017, Polydoropoulou et al., 2021).

Finally, considering the more limited specification for the transit sample, initial transit ICLV specifications included the Pro-Sharing Economy latent variable; however, it was not identified even when the Structural Equation Model was estimated independently of the discrete choice model. Consequently, the only latent variable identified for transit commuters is COVID Comfort.

# 6 Implications
## 6.1 Microtransit Demand and Curb-to-curb Attribute Elasticities

The curb-to-curb attributes involving out-of-vehicle travel time were only statistically significant in the car commuter ICLV. In contrast, transit commuters were unaffected by the walking and waiting time. We hypothesize that this is due to transit commuters already experiencing these attributes for their current commutes. Therefore, when trying to attract car commuters to microtransit to promote sustainability, attention must be paid to the effort needed to access the service in terms of expected walking and waiting time.

One strategy is to decrease waiting and walking times and to increase the minimum reservation time to facilitate better routing. To better explore such scenarios and the relative importance of microtransit attributes, we derive attribute elasticities. **Table** *7* shows the elasticities at the mean of variables, which were calculated using Equation 5 (Train, 2009). $P_i$ is the probability of alternative i, $\beta_{x,i}$ is the coefficient of attribute $x$ and alternative i, and $x_i$ is the average of the explanatory variable. These elasticities reflect the percent change in demand for the alternative as a function of a unit percent change in the attribute. We note that most elasticities are inelastic, in the range of 4-76% change in demand for the Microtransit options. As expected from the model analysis, reservation time has a lower elasticity than in-vehicle, waiting, and walking time. In comparing the commuter groups, elasticities are in a comparable range for the sedan option, with a greater gap for the van microtransit option. Clearly, drivers are sensitive to more attributes and display significant aversion to access/walking time, while in-vehicle travel duration elasticity even exceeds unity for the van option.

**Table 7 Elasticities and Difference between Commuter Groups**

| Alternative | Variable | Elasticities | | Difference (Car-Transit) |
| --- | --- | --- | --- | --- |
| | | Car Commuters | Transit Commuters | |
| Status Quo | Cost | -0.04 | -0.30 | 0.26 |
| Status Quo | TT | -0.41 | -0.62 | 0.21 |
| Status Quo | COVID | -0.94 | -0.20 | -0.74 |
| Status Quo | Sharing (Indirect effect) | -0.83 | NA | |
| MT-S | Cost | -0.59* | -0.39 | -0.2 |
| MT-S | TT | -0.77 | -0.71 | -0.06 |
| MT-S | Reservation Time | -0.04 | -0.07 | 0.03 |
| MT-S | Wait | -0.13 | NS | |
| MT-S | Walk | -0.13 | NS | |
| MT-V | Cost | -0.49* | -0.62 | 0.13 |
| MT-V | TT | -1.26 | -0.64 | -0.62 |
| MT-V | Reservation Time | -0.12 | -0.19 | 0.07 |
| MT-V | Wait | -0.31 | NS | |
| MT-V | Walk | -0.39 | NS | |

(*) Cost parameters are for car commuters with commutes > 65 minutes

Operators can use these insights in several ways. Microtransit operators may unlock efficiency gains and reductions of passenger wait times by knowing the demand for rides well in advance. Indeed, the smaller elasticity suggests that increasing minimum reservation time would not be as consequential for the likelihood to opt for the microtransit alternatives as increasing walk and wait times. Thereby, the elasticity findings suggest an opportunity to extend reservation times to obtain more favorable walking and waiting performance as a means to attract drivers to

$$E = (1 - P_i)\beta_{x,i}X_i \qquad (5)$$

the curb-to-curb mobility options. Similar to Alonso-González et al. (2020a), this reduction in travel time plays a prominent role in determining the likelihood of choosing microtransit. To further contextualize, Alonso-Mora et al. (2017) simulate scenarios with maximum waiting times of less than 7 minutes; however, this was in the highly-dense area of Manhattan, New York where high levels of demand and the road network topology allow this. Therefore, for success in less dense areas, a large vehicle fleet size is another strategy to reduce wait and walking times.

For transit commuters, much of the focus for microtransit operators will be on cost and travel time as these commuters did not exhibit significant sensitivity to waiting and walking times. One attribute that was only significant in a single instance was the sheltered boarding location. While this may be a prominent feature for public transit, it may not be a worthwhile investment in this context, where other curb-to-curb attributes play a greater role in shaping initial demand for microtransit.

### 6.2 Different Perceptions for Drivers and Transit Commuters: Status Quo Effects

When considering the latent variables identified in the ICLVs, the lack of Pro-Sharing Economy in the transit commuter group is intriguing. It was expected that Pro-Sharing Economy attitudes would be identified in the transit group since this mode embodies shared mobility, yet our modeling did not support this. Additionally, COVID Comfort is only weakly significant ($0.10 < $ p-value $ < 0.05$). Taken together, the latent variable results suggest that the transit users in this sample are likely captive users (Etzioni et al., 2020). Indeed, the analysis of smartcard usage conducted before the pandemic shows that heavy users of transit in Israel are more likely to be regarded as captive with fewer mobility options—pupils, students, seniors, low income—while the modal split for the Tel Aviv metropolitan region is around 80/20 for car and transit respectively (Benenson et al., 2019, Etzioni et al., 2021).

Instead, both latent variables are strongly significant in the car ICLV. Because Pro-Sharing Economy is mainly determined by experience with sharing economy services like Uber and Airbnb, we hypothesized that knowledge and familiarity with these types of services would lower the risk perceptions related to COVID-19. What is more, operators have taken significant and public measures to increase patrons' safety, which may have contributed to indirectly shaping virus exposure concerns in the context of hypothetical microtransit alternatives. Therefore, unlike transit commuters, we do not conclude that car commuters are captive to their status quo. Similarly, the elasticity of the COVID-19 comfort variable is much larger among car commuters. We interpret this strong effect to reflect greater adaptiveness of drivers in response to COVID-19. It is reasonable to assume that those who rely on private vehicles have greater ease in adjusting ridership to reduce the risk of viral exposure.

## 7  Conclusions and Considerations for Future Research

Microtransit with rider pooling may generate mobility system benefits, chief among them being the VMT reductions unlocked if enough trips are pooled. The demand for microtransit, especially with a curb-to-curb service offering, is not fully understood. Specifically, it is challenging to promote the adoption of microtransit given that the service attributes lie at the intersection between door-to-door (private driving) or on-demand (ride-hailing) mobility and scheduled transit. This implies that current mode experiences are likely to shape the perception of microtransit attributes. Such insight becomes critical to consider given the need for microtransit to attract not only transit users to ensure effective VMT and congestion reduction. In this study we developed an SC survey to identify how commuters perceive microtransit including curb-to-curb attributes. Specifically, the experiment included tailored designs for car and transit commuters. Utilizing a pivoted design with the status quo alternative, we identified how sensitive commuters are to a sedan (MT-S) and a van option (MT-V) and their curb-to-curb attributes such as walking and waiting times at a designated boarding location. Additionally, we included attributes that better represent the scheduling component of microtransit, where advanced planning and amenities are key attributes. Specifically, we included novel attributes for minimum reservation time before boarding and a sheltered boarding location. The results reveal differences among commuter groups. While car commuters were sensitive to walking and waiting time, transit commuters were not. Minimum reservation time significantly affected the utility of the microtransit alternatives; however, the elasticities show that in- and out-of-vehicle travel time have larger effects. Among these novel attributes, the sheltered boarding location had no significant effect on the utility of the shared modes except for MT-V for transit commuters.

This analysis took place after pandemic lockdown periods, and several questions were designed to measure COVID-19 risk and comfort to quantify the potential impacts. The latent variable portion of the ICLV reveals that COVID Comfort affects utility for car commuters far more than transit commuters. Furthermore, the Pro-Sharing Economy latent variable was not even identified for transit commuters, although public transit is defined by sharing. We take these results to suggest that transit commuters are very likely captive, reflecting a limited sensitivity to the latent constructs, and focus on the main attribute effects. Instead car commuters appear to be more responsive to the latent effects, suggesting more adaptiveness to changing circumstances for both shared space and COVID-exposure. On the whole, this study gives added understanding about nuances in COVID-19 impacts on mobility choices, showing that the risk perceptions across commuter groups are far from uniform.

Based on these results, operators of microtransit services must consider several strategies to attract riders. These strategies ought to be differentiated by commuter groups as they show major divergence in attribute and latent variable effects. Looking more carefully at the trade-offs gives valuable insights. For instance, in the sedan setting, car commuters are not averse to adding more passengers. Instead in the van context, additional passengers come at the cost of highly coveted travel time, suggesting that car commuters need to be compensated by careful calibration of cost-fare tradeoffs.

There are limitations to this study that should be noted. The sampling for this web-based survey may not represent the entire commuter population, especially digitally challenged citizens. Secondly, the survey and modeling were done separately for the commuter groups. Given the group-specific tailoring of the experiment the modeling needed to be separated, while a joint analysis may reveal other phenomena. The timing of the study in the context of the pandemic also introduces a caveat of not representing the rapidly evolving circumstances. A later administration would likely reflect different decision given the proliferation of new viral variants, and the rapid immunization campaign in Israel.

Future research should consider probing the lack of significant cost variables for the car commuters. The choice experiment design attempted used here aimed at realism by only providing tasks where microtransit costs were lower than the status quo, possibly at the cost of capturing the full cost sensitivity. One possibility to study both the dynamics of the pandemic, and the effect of pricing variations, would be to incorporate gamification into the mode choice experiment (Klein and Ben-Elia, 2016). A dynamic choice experiment that updates mode attributes based on user responses would improve understanding of how cost affects mode choice. For example, continually decreasing the cost of the microtransit alternatives until the respondent switches from the status quo may result in an improved understanding of the cost tradeoffs. Additionally, the context surrounding the COVID-19 pandemic is continuously evolving as information and vaccination rates are changing. Future research should incorporate broader pandemic effects, such as social distancing measures, and real-time information about pandemic indicators, to parse the effects on microtransit adoption and competition among modes with different levels of sharing and exposure.

**Acknowledgements**
This research was partially funded by the Chief Scientist Office at the Israeli Ministry of Transport and Road Safety and The Transportation Innovation Institute at Tel Aviv University; and received partial funding from the U.S. National Science Foundation (NSF) Career grant No. 1847537. An earlier version of this paper was submitted for presentation to the 3rd Annual Bridging Transportation Researchers (BRT) in 2021, and for presentation to the 2022 edition of Transportation Research Board (TRB) annual conference.

**Conflict of interest**
On behalf of all authors, the corresponding author states that there is no conflict of interest.

**Authors' contribution**
Conceptualization: Ben-Elia, Study Design: Ben-Elia, Stathopoulos, Soria; Methodology: Ben-Elia, Stathopoulos, Soria; Data curation: Soria; Investigation: Soria and Etzioni; Formal analysis: Soria; Writing - original draft preparation: Soria; Writing - review and editing: Ben-Elia, Stathopoulos, Soria, Etzioni, Shiftan; Software And Visualization: Soria; Funding acquisition: Ben-Elia and Shiftan; Resources: Ben-Elia and Stathopoulos; Supervision: Ben-Elia and Stathopoulos

CAUSSADE, S., DE DIOS ORTÚZAR, J., RIZZI, L. I. & HENSHER, D. A. 2005. Assessing the influence of design dimensions on stated choice experiment estimates. *Transportation research part B: Methodological,* 39**,** 621-640.
CHAVIS, C. & GAYAH, V. V. 2017. Development of a Mode Choice Model for General Purpose Flexible-Route Transit Systems. *Transportation Research Record: Journal of the Transportation Research Board***,** 133-141.
CHEN, X., ZHENG, H., WANG, Z. & CHEN, X. 2018. Exploring impacts of on-demand ridesplitting on mobility via real-world ridesourcing data and questionnaires. *Transportation*.
CHICAGO METROPOLITAN AGENCY FOR PLANNING 2019. New Data Allows an Initial Look at Ride Hailing in Chicago.
CHOICEMETRICS 2012. Ngene 1.1. 1 user manual & reference guide. *Sydney, Australia*.
CLARK, B. Y. & BROWN, A. 2021. What does ride-hailing mean for parking? Associations between on-street parking occupancy and ride-hail trips in Seattle. *Case Studies on Transport Policy,* 9**,** 775-783.
CLEWLOW, R. R. & MISHRA, G. S. 2017. Disruptive transportation: The adoption, utilization, and impacts of ride-hailing in the United States. *University of California, Davis, Institute of Transportation Studies, Davis, CA, Research Report UCD-ITS-RR-17-07*.
CORREA, D., XIE, K. & OZBAY, K. Exploring the taxi and Uber demand in New York City: An empirical analysis and spatial modeling. 2017 2017.
DALY, A. 2010. Cost damping in travel demand models: Report of a study for the department for transport.
DAS, S., BORUAH, A., BANERJEE, A., RAONIAR, R., NAMA, S. & MAURYA, A. K. 2021. Impact of COVID-19: A radical modal shift from public to private transport mode. *Transport Policy,* 109**,** 1-11.
DE VOS, J. 2020. The effect of COVID-19 and subsequent social distancing on travel behavior. *Transportation Research Interdisciplinary Perspectives,* 5**,** 100121.
DONG, X. 2020. Trade Uber for the bus? An investigation of individual willingness to use ride-hail versus transit. *Journal of the American Planning Association,* 86**,** 222-235.
DUARTE, L. 2020. Last ride for Chicago's taxis? Ride-sharing pandemic taking its toll. *WGN*.
ERHARDT, G. D., MUCCI, R. A., COOPER, D., SANA, B., CHEN, M. & CASTIGLIONE, J. 2021. Do transportation network companies increase or decrease transit ridership? Empirical evidence from San Francisco. *Transportation***,** 1-30.
ERHARDT, G. D., ROY, S., COOPER, D., SANA, B., CHEN, M. & CASTIGLIONE, J. 2019. Do transportation network companies decrease or increase congestion? *Science advances,* 5**,** eaau2670.
ETZIONI, S., DAZIANO, R. A., BEN-ELIA, E. & SHIFTAN, Y. 2021. Preferences for shared automated vehicles: A hybrid latent class modeling approach. *Transportation Research Part C: Emerging Technologies,* 125**,** 103013.
ETZIONI, S., HAMADNEH, J., ELVARSSON, A. B., ESZTERGÁR-KISS, D., DJUKANOVIC, M., NEOPHYTOU, S. N., SODNIK, J., POLYDOROPOULOU, A., TSOUROS, I. & PRONELLO, C. 2020. Modeling cross-national differences in automated vehicle acceptance. *Sustainability,* 12**,** 9765.
FOWLER, J. H. 2006. Altruism and turnout. *The Journal of Politics,* 68**,** 674-683.
Page **27** of **31**